\documentclass{iaus}
\usepackage{graphicx}

\title[LBG Evolution from z=5 to 3] 
{Luminosity Dependent Evolution of Lyman Break Galaxies from redshift 5 to 3}

\author[Iwata et al.]   
{Iwata I.$^1$,%
Ohta K.$^2$, 
Tamura N.$^3$, 
Akiyama M.$^3$, 
Aoki K.$^3$, 
Ando M.$^2$, 
Kiuchi G.$^2$, 
\and
Sawicki M.$^{2,4}$}

\affiliation{$^1$Okayama Astrophysical Observatory, 
National Astronomical Observatory of Japan, 
Kamogata, Asakuchi, Okayama, Japan 719-0232 
\break email: iwata@oao.nao.ac.jp \\[\affilskip]
$^2$Department of Astronomy, Graduate School of Science, 
Kyoto University, Sakyo-ku, Kyoto, Japan 606-8502\\[\affilskip]
$^3$Subaru Telescope, 
National Astronomical Observatory of Japan, 
650 North A`ohoku Place, Hilo, HI 96720\\[\affilskip]
$^4$Physics Department, University of California, Santa Barbara, 
CA 93106, USA
}

\pubyear{2006}
\volume{235}  
\pagerange{1--2}
\date{?? and in revised form ??}
\jname{Galaxy Evolution across the Hubble Time}
\editors{F.Combes \& J. Palous, eds.}
\begin{document}

\maketitle

\begin{abstract}
The development of large ground-based telescopes and 
sensitive large format detectors, 
as well as the develepment of various techniques for the 
selection of high-z galaxies enabled us to construct 
large samples of galaxies in the early universe, 
as reported in the many contributions in this proceedings. 
The next major step for the comprehensive understanding 
of the galaxy evolution would be to explore the relationship 
of galaxies selected with different criteria at different epochs 
and find links between them.
In this contribution we present the properties of Lyman break galaxies 
(LBGs) at $z \sim 5$ obtained by deep and wide blank field surveys, 
and through the comparison with samples at lower redshift ranges we 
discuss the evolution of star-forming galaxies in the early universe.
\end{abstract}


Our $z \sim 5$ LBG sample is based on the deep and wide surveys for the 
two independent blank fields (the region including the Hubble Deep 
Field - North and the J0053+1234) obtained with the Suprime-Cam 
attached to the 8.2m Subaru Telescope. The total effective area after masking  
bright objects is 1,300 arcmin$^2$, and deep $V$, $I_c$ and $z'$-band 
imaging enabled us to securely select $V$-dropout objects down to 
$z'_\mathrm{AB}=26.5$ mag (for the HDF-N region) or 25.5 mag (for the 
J0053+1234 region). The number of LBG candidates in our sample is 
850. It should be emphasized that the area coverage our survey 
is more than 100 times wider than 
the ACS field of the Hubble Ultra Deep Field and more than 4 times 
wider than the total area covered by the GOODS, and this wide field 
coverage has a crucial importance for reliable determination of the 
abundance of luminous objects. 
So our survey are able to explore both bright and faint parts of 
the LF reliably. The redshift of a part of our LBG candidates 
have been spectroscopically determined (\cite[Ando et al. 2004]{Ando04}), 
and the validity of our color selection criteria have been confirmed.

\begin{figure}
 \resizebox{9cm}{!}{\includegraphics{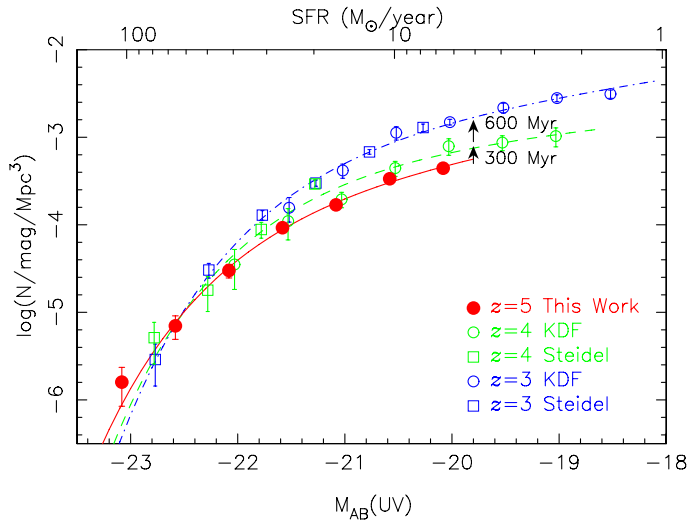}}
  \caption{UV luminosity function of LBGs from $z=5$ to 3.}\label{fig:LF}
\end{figure}

In figure~\ref{fig:LF} we show the UV luminosity function (LF) of 
LBGs at $z \sim 5$ derived statistically from our sample with 
filled circles and a solid line (Iwata et al. submitted to MNRAS). 
In this figure we also 
show the UVLF of LBGs at $z\sim 4$ and 3, taken from the 
deep survey by \cite{KDF2}. We found that in the luminous end 
of the UV LF there is no significant evolution from $z\sim5$ to 
3 ($\approx$1 Gyr), while in the fainter part, the gradual 
increase of number density is observed. This trend has been 
suggested in \cite{Iwata03} but now it becomes much 
significant thanks to the improvement of the data. 
This clear contrast 
in the UV LF suggests that the evolution of the LBGs is 
{\it differential} depending on the UV luminosity.

Our follow-up studies on the LBGs at $z \sim 5$ are on-going, 
and there are several intriguing results supporting the differential 
evolution of the LBGs; (1) in optical spectroscopy we found 
that in the UV luminous LBGs at $z \sim 5$ 
there is no or little Lyman$\alpha$ emission, and 
Lyman$\alpha$ equivalent widths depends on UV luminosity 
for LBGs at $z=5$--6 (\cite[Ando et al. 2006]{Ando06}).
These trends suggest either the heavier dust attenuation 
in the luminous LBGs and/or the existence of massive neutral 
gas surrounding them. 
(2) Both restframe UV colors and UV-to-optical colors 
(obtained by the cross identification with 
Spitzer space telescope / IRAC images) 
show luminosity dependence; luminous objects 
tend to be red (\cite[Iwata et al. 2006]{Iwata06}).
(3) From the SED fitting for LBGs at $z \sim 5$ with detections 
in the $K'$-band and IRAC 3--5 $\mu$m images, 
stellar masses as large as $>10^{10} M_\odot$ have been estimated. 
It suggests a remarkable star formation in the first 1 Gyr of the 
universe. 
(4) Stronger clustering has been detected for the luminous 
LBGs at $z \sim 5$ than that for fainter ones, implying that 
luminous LBGs reside in massive dark matter haloes.
From these findings we suggest that the evolution of star-forming 
galaxies in the first 2 Gyr of the universe could be well described 
with the biased evolution scenario:  
a galaxy population hosted by massive dark haloes start active 
star formation preferentially at early time of the universe, 
while less massive galaxies increase their number density later. 
To understand the origin of this differential evolution would 
be an important clue to clarify a star-formation process 
in the early universe.

\end{document}